\documentclass[a4paper,fleqn]{cas-dc}

\usepackage[numbers]{natbib}

\usepackage{amsmath,amsfonts,amssymb}
\usepackage{graphicx}
\usepackage{xcolor}
\usepackage{booktabs}
\usepackage{siunitx}
\usepackage{listings}
\usepackage{tikz}
\usepackage{todonotes}
\usepackage{pgfplots}
\pgfplotsset{compat=1.18}
\usetikzlibrary{arrows.meta,positioning,shapes.geometric,shapes.multipart}
\usepackage[ruled,vlined]{algorithm2e}

\begin{document}
\let\WriteBookmarks\relax
\def\floatpagepagefraction{1}
\def\textpagefraction{.001}

\shorttitle{$\lambda$PIC: A callback-centric PIC framework}

\shortauthors{X.S. Geng et~al.}

\title[mode=title]{$\lambda$PIC:\ A callback-centric particle-in-cell framework}

\author[1]{Xuesong Geng}
\cormark[1]
\ead{xsgeng@siom.ac.cn}
\credit{Conceptualization, Methodology, Software, Fundings acquisition, Writing-original draft, Visualization}

\author[1]{Yunwei Cui}
\credit{Methodology, Software}

\author[1]{Lingang Zhang}
\credit{Methodology, Software, Funding acquisition}

\author[1]{Liangliang Ji}
\cormark[1]
\ead{jill@siom.ac.cn}
\credit{Supervision, Writing-review \& editing, Funding acquisition}

\affiliation[1]{organization={State Key Laboratory of Ultra-intense laser Science and Technology, Shanghai Institute of Optics and Fine Mechanics (SIOM), \\Chinese Academy of Sciences (CAS)},
  city={Shanghai},
  postcode={201800},
  country={China}}

\cortext[1]{Corresponding author}

\begin{abstract}
  We present $\lambda$PIC, a Python-based electromagnetic particle-in-cell framework built around a callback-centric architecture. Existing PIC codes typically tie high performance to static, pre-compiled timestep loops, hindering implementation of custom physics, diagnostics, or output logic. $\lambda$PIC breaks this coupling by exposing every stage of the loop as a named stage (hook), permitting attaching arbitrary Python functions that operate on the full simulation state, enabling custom algorithms and in-situ analysis without modifying the core algorithms. Under this flexible framework, performance-critical kernels are written in C extensions and Numba, fields and particles are stored in NumPy arrays, and MPI parallelism is paired with graph partitioning to support dynamic load balancing and non-rectangular domains. Although $\lambda$PIC is designed as general-purpose, it has special focus on intense laser-plasma interactions. Future work will extend the framework to GPU acceleration and additional physics modules including implicit solvers and nuclear physics.
\end{abstract}

\begin{highlights}
  \item Callback-centric architecture lets users attach arbitrary Python functions for custom physics and diagnostics.
  \item Performance-critical kernels are written in C and Numba with data exposed as NumPy arrays.
  \item Graph-based domain decomposition enables dynamic load balancing and non-rectangular geometries.
\end{highlights}


\maketitle

\section{Introduction}\label{sec:intro}

Particle-in-cell (PIC) simulation has been the dominant method for kinetic plasma physics for more than five decades, providing self-consistent solutions to the Vlasov--Maxwell system through the motion of representative macro-particles on a computational grid~\cite{c.k.birdsallPlasmaPhysicsComputer1991,hockneyComputerSimulationUsing2021}.
During this period, a rich ecosystem of open-source frameworks has matured to support research in laser-plasma interaction, astrophysical plasmas, and accelerator physics.

Each framework makes different trade-offs among execution speed, flexibility, and development productivity.
Most frameworks use a pre-compiled computational core run from static input files~\cite{arberContemporaryParticleincellApproach2015,fonsecaOSIRISThreeDimensionalFully2002,pukhovThreedimensionalElectromagneticRelativistic1999,derouillatSmileiCollaborativeOpensource2018}.
Simulation parameters and physics modules are selected at run time, but the available operations are fixed at compile time; adding new physics generally requires modifying the core source and recompiling.

More flexible frameworks provide a Python frontend that coordinates a pre-compiled kernel library, achieving performance through just-in-time compilation or native bindings~\cite{brogrenpiPICFrameworkModular2026,vayBLASTWarpXWarpx26062026,leheSpectralQuasicylindricalDispersionfree2016a}.
Some of these frameworks allow user-defined functions to be injected at specific points in the compiled timestep loop through a callback or handler mechanism.
However, the callback locations are fixed by the compiled backend and the core solver logic remains hard-coded.
Operations like output, laser injection, and standard diagnostics are embedded in the backend source, with user callbacks serving as optional extensions to this predefined loop.
Moreover, existing callback mechanisms impose structural restrictions on data access,
often requiring state to be retrieved through indirect APIs or compiled interfaces.
These restrictions hinder global state operations, rapid prototyping, and purely Python-based extensions.

$\lambda$PIC departs from this pattern by making callbacks the primary extension mechanism.
The main timestep loop contains no hard-wired output, diagnostic, or injection logic.
Every operation is registered as a callback.
User-defined Python functions attach to named stages and receive the full simulation state, enabling direct access to all field and particle arrays as NumPy objects without compiled extensions or domain-specific APIs.
This architecture rests on three design choices.
First, the callback system exposes named stages throughout the main timestep loop.
Because callbacks operate on the full state, they can inject custom physics, analyze intermediate quantities in situ, or extract custom diagnostics.
This enables rapid prototyping and clean extensions that would otherwise require changes to the core code.
Second, a Python/Numba/C-extension hybrid achieves high performance through just-in-time compilation and dedicated C kernels.
A hybrid MPI-OpenMP model parallelizes the work over patches, while the Python layer handles coordination and user extensions.
Third, graph partitioning replaces the traditional Cartesian patch decomposition, enabling dynamic load balancing and non-rectangular domain shapes.

The remainder of this paper is organized as follows.
Section~\ref{sec:architecture} describes the callback-centric architecture, the patch-based data layout, and the domain decomposition.
Section~\ref{sec:numerical} summarizes the numerical methods.
Section~\ref{sec:verification} presents code verification against standard benchmark problems.
Section~\ref{sec:scaling} reports strong and weak scaling results and dynamic load balancing.
Section~\ref{sec:applications} showcases $\lambda$PIC on specific problems.
Section~\ref{sec:conclusions} concludes the paper and outlines directions for future development.
$\lambda$PIC is open-source and publicly available at \url{https://github.com/xsgeng/lambdapic}.

\section{Architecture}\label{sec:architecture}

\subsection{Class hierarchy and data model}\label{sec:class-hierarchy}

\begin{figure*}[t]
  \centering
  \includegraphics[width=\textwidth]{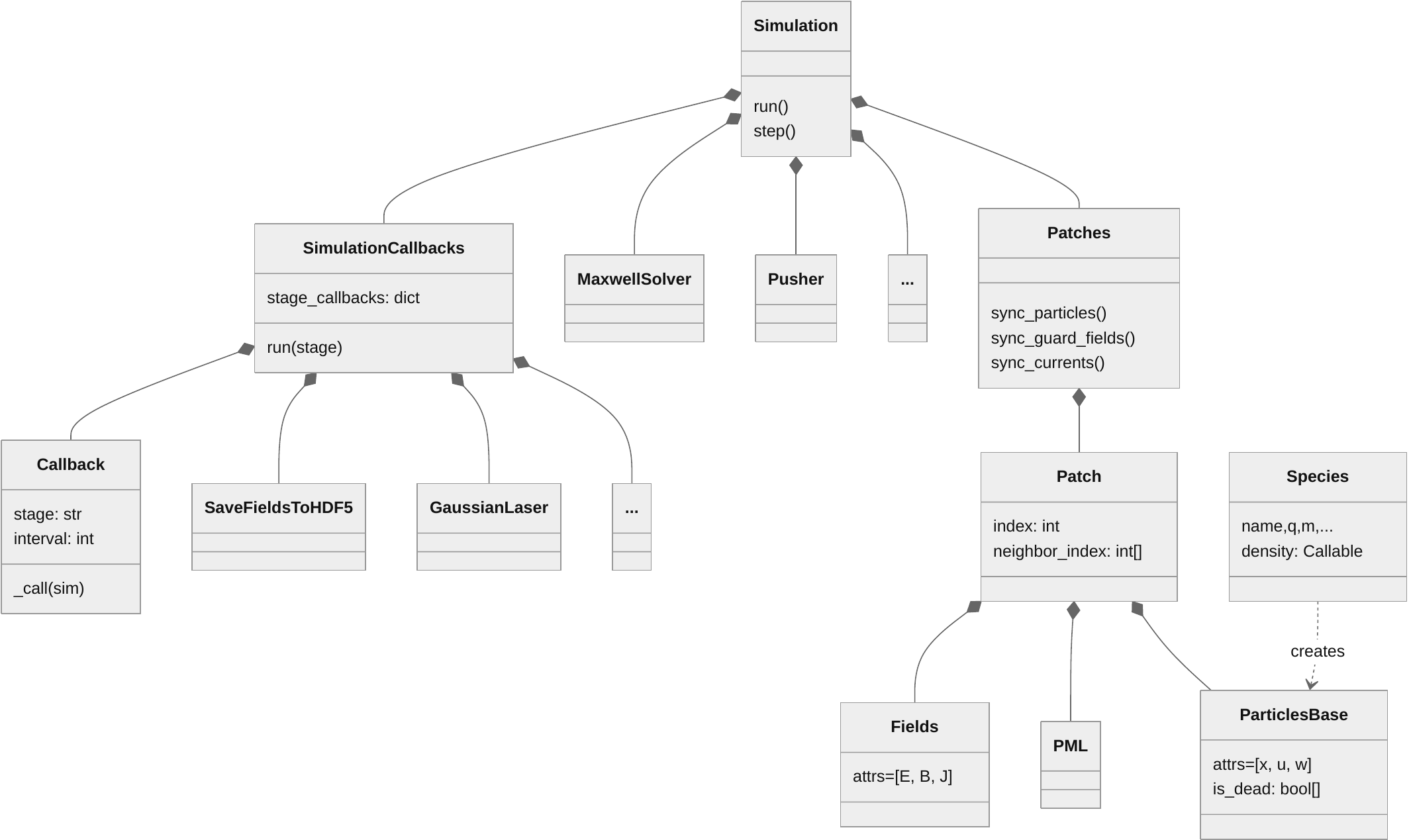}
  \caption{
    Class diagram of the $\lambda$PIC core architecture.
    A top-level \texttt{Simulation} object composes the \texttt{Patches} container, physics modules (\texttt{MaxwellSolver} and \texttt{Pusher} for example), and the callback registry (\texttt{SaveFieldsToHDF5} and \texttt{GaussianLaser} for example).
    \texttt{Patches} contains \texttt{Patch} instances, each holding field data, PML boundaries, and particle data.
    Not all components and relationships are shown for clarity.
  }
  \label{fig:class-diagram}
\end{figure*}

The $\lambda$PIC architecture (Fig.~\ref{fig:class-diagram}) is orchestrated by a top-level \texttt{Simulation} object that composes the \texttt{Patches} container, physics modules, and the callback registry.
The \texttt{Patches} container aggregates spatial subdomains and serves as the shared data layer for all modules.
Rather than embedding solver logic inside a monolithic class, each module holds a reference to \texttt{Patches} and operates on the field and particle arrays stored inside.
In addition to the shared \texttt{Patches} reference, each module manages its own internal data independently.
This loose coupling simplifies testing, because any module can be exercised in isolation by constructing a minimal \texttt{Patches} instance.

The spatial domain is decomposed into a Cartesian array of rectangular subdomains called \emph{patches}.
The Patches container manages all patches that reside on a single MPI rank; each patch is represented by a Patch instance.
Each patch stores its own electromagnetic fields instance, a list of particle arrays (one per species), and optional perfectly-matched layer (PML) boundaries~\cite{berengerPerfectlyMatchedLayer1994}.
The fields data are allocated in the Yee arrays~\cite{yeeNumericalSolutionInitial1966} of $\mathbf{E}$, $\mathbf{B}$, $\mathbf{J}$, and $\rho$ with a configurable number of guard cells.

Particles are stored in a Structure-of-Arrays (SoA) layout, a choice that is now standard in high-performance PIC codes because it enables contiguous memory access and efficient SIMD vectorization~\cite{homannSoAxGenericStructure2018,barsamianEfficientDataLayouts2018,gruberLLAMALowlevelAbstraction2023}.
For every macro-particle, positions $(x,y,z)$, normalized momenta $(u_x,u_y,u_z)$, inverse Lorentz factor $\gamma^{-1}$, weights $w$, a death flag \texttt{is\_dead}, and a bit-packed unique identifier are held as one-dimensional arrays.
The macro-particle class has an \texttt{attrs} list that declares which arrays participate in initialization, memory management, migration, and synchronization, making the data model self-describing.

Both the fields and particle arrays are stored in NumPy arrays for best compatibility for arbitrary user operations.
And both classes expose an \texttt{attrs} list that enumerates the arrays they own.
Any array named in \texttt{attrs} is automatically allocated, resized and synchronized.
A user can therefore attach arbitrary per-particle or per-cell data (for example, quantum nonlinearity parameter $\chi$, optical depth $\tau$ and spin) by appending the attribute name to \texttt{attrs}.
The new arrays then participate transparently in intra-patch boundary exchange, inter-rank MPI transfer, and dynamic load balancing without any modification to the core solver code.
The \texttt{Fields} class supports the same mechanism.
It should be noted that the evolution of the added attributes needs corresponding particle or field algorithms.

Deleting a macro-particle is a frequent event in PIC simulations, because particles leave the domain, are absorbed at boundaries, or are removed by QED processes.
Rather than immediately shrinking the host array, which would require an $O(N)$ compacting copy at every deletion, or resorting to frame-based pools as in some GPU-native codes~\cite{leeAccelerationParticleincellCode2025}, $\lambda$PIC marks the macro-particle as dead by setting its \texttt{is\_dead} flag to \texttt{True}.
The slot remains allocated and is eligible for reuse.
During inter-patch particle migration, incoming particles preferentially fill dead slots before the array is grown, so in regimes where the outgoing and incoming fluxes balance, no memory reallocation occurs at all.
The \texttt{prune()} method compacts the alive particles to the front of each array in a single vectorized \texttt{argsort}, resizes the arrays down to $\text{alive} \times (1 + \text{extra\_buff})$, and leaves a small buffer for future insertions.
This lazy-deletion strategy reduces memory overhead, preserves the contiguity required for Numba-vectorized kernels, and keeps the per-particle memory footprint minimal because only one extra boolean array is required, in contrast to merging-based population control sometimes used in QED-cascade simulations~\cite{vranicParticleMergingAlgorithm2015}.

Species are defined by the species dataclass.
Concrete species such as \texttt{Electron}, \texttt{Positron}, \texttt{Proton}, and \texttt{Photon} set default charge and mass.
Each species carries a user-supplied number density profile, a macro-particle-per-cell count, and momentum-distribution functions.
The density and particles-per-cell (PPC) functions are Just-In-Time (JIT) compiled with Numba~\cite{lamNumbaLLVMbasedPython2015} via a small reflection layer that accepts Python callables, constants, or already-compiled numba dispatcher.
When a species is added to the simulation it calls the particle creation method, which acts as a factory returning the appropriate particle class.

\subsection{Callback system}

A distinguishing feature of $\lambda$PIC is that the main timestep loop contains no hard-wired output, diagnostic, or injection logic.
The framework is callback-centric so that researchers can prototype new physics, perform complex in-situ diagnostics on intermediate quantities, and carry out inline analysis without post-processing, all by injecting user-defined logic at predefined stages.
Every operation, including data I/O, plotting, checkpointing, laser injection, and arbitrary user-defined logic that may inspect, analyze, or modify the simulation state, is registered as a callback.
The callback system is therefore not an afterthought but the primary mechanism by which the framework is extended.

The simulation pipeline exposes many predefined callback stages ranging from one-time initialization to per-timestep events and physics stages such as when the Maxwell solver completes or the particle pusher completes.
$\lambda$PIC has built-in callbacks such as \texttt{SaveFieldsToHDF5}, \texttt{PlotFields}, \texttt{GaussianLaser}, \textit{etc.}\ for ease of use.

Two registration mechanisms are provided.
For simple functions the \texttt{@callback} decorator attaches stage and interval metadata.
The decorator wraps the function with interval logic (integer step counts, floating-point simulation-time intervals, or arbitrary predicates), MPI-aware timing, and a barrier so that all ranks remain synchronized.
For stateful callbacks the \texttt{Callback} base class is subclassed; the derived class sets the stage and interval in its constructor and implements its own callable.

\subsection{Performance modules}

The field update and species initialization are JIT-compiled with Numba~\cite{lamNumbaLLVMbasedPython2015}.
This preserves the ability to pass arbitrary user-defined density profiles or momentum distributions without sacrificing performance.
The critical kernels (field interpolation, current deposition, pushers, particle synchronization and other per-particle operations) are implemented in C.
By keeping the inner loops in compiled code, the framework eliminates the per-particle Python overhead while user callbacks retain full control of the high-level orchestration.

When no callbacks are registered between the intermediate stages of the particle update, $\lambda$PIC activates the unified pusher.
It fuses the position push, field interpolation, momentum push, and current deposition into a single C loop over particles.
This reduces memory-bandwidth pressure and avoids the repeated Python-level dispatch overhead.

Inside each MPI rank the workload is further divided among OpenMP threads at the patch level.
All compute kernels iterate over the local patch list in an outer parallel loop.
In C extensions this is implemented with OpenMP; in Numba-compiled routines this uses parallel range over patches.
Each patch holds independent field and particle arrays, so the threads operate on separate memory regions and need not synchronize during the compute phase.

\subsection{Domain decomposition and graph partitioning}\label{sec:domain-decomp}

By default $\lambda$PIC creates a regular Cartesian array of patches.
Flexible, workload-aware load balancing is achieved through graph partitioning rather than space-filling curves~\cite{germaschewskiPlasmaSimulationCode2016,derouillatSmileiCollaborativeOpensource2018}.
Graph partitioning is already well established for load balancing unstructured-mesh simulations~\cite{diamondDynamicLoadBalancing2018,bettencourtEMPIREPICPerformancePortable2021,diamondPUMIPicMeshbasedApproach2021}.
It is applied to the patch grid of $\lambda$PIC.
Each patch stores the global indices of its face, edge, and corner neighbors.
In two dimensions there are 8 neighbors; in three dimensions there are 26 neighbors.
Periodic boundary conditions are handled by linking the boundary indices to the opposite side of the domain.
Based on the neighbor tables, guard-cell fields and particles crossing patch boundaries are synchronized via shared-memory or MPI routines.

\begin{figure}[]
  \centering
  \includegraphics[width=\columnwidth]{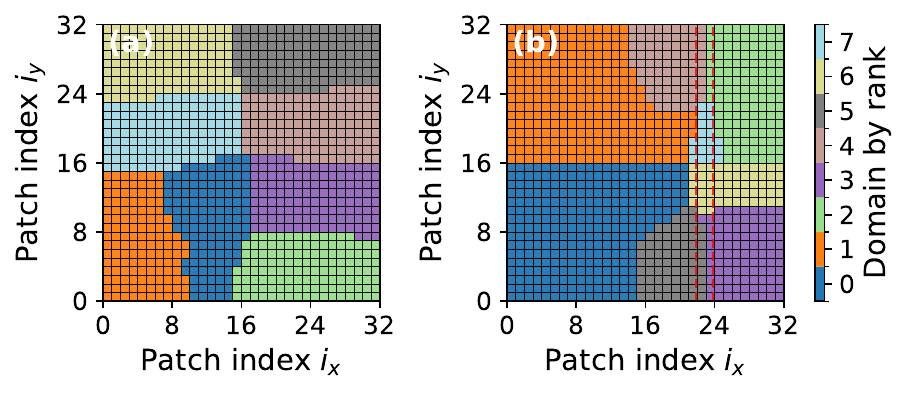}
  \caption{
    Graph-based domain decomposition for a $32 \times 32$ patch grid on eight MPI ranks. 
    (a) Uniform density and (b) plasma slab density profile with boundaries denoted by red-dashed line.
    Axes are patch indices $i_x$ and $i_y$; color of the patch denotes the MPI rank.
  }
  \label{fig:metis-partition}
\end{figure}

Unlike wiring the patches with space filling curves,
patches are decomposed into an arbitrary topology via graph partitioning (METIS~\cite{karypisFastHighQuality1998} in this implementation).
The simulation domain can be divided into non-rectangular domains according to the weights of patches, as shown in Fig.~\ref{fig:metis-partition}.
When every patch carries the same vertex weight the decomposition is a regular patch-like pattern (Fig.~\ref{fig:metis-partition}a).
If the load is spatially non-uniform, for example in laser-target interactions, high-load patches in the target region are assigned to some ranks and other ranks are assigned more vacuum patches (Fig.~\ref{fig:metis-partition}b).
It should be noted that, due to the shared-memory nature of the patch-parallelism, the number of vertices is small (equals the number of non-uniform memory access (NUMA) nodes), and the re-partitioning cost is low.

The graph is constructed directly from the neighbor arrays. Each patch is a vertex, and an edge connects adjacent patches.
The partitioner is invoked with the contiguous option so that every rank owns a contiguous region of patches, which reduces the communication volume during boundary exchange.
This approach naturally supports non-rectangular geometries; a proof-of-concept demonstration with an annular domain is presented in Section~\ref{sec:annular}.

A rank-stable mapping minimizes data movement during dynamic rebalancing.
When the partitioner returns a new partition, an assignment algorithm matches the new rank IDs to the old ones by maximizing the overlap of patch sets between the old and new distributions.
Patches that do not change owner therefore require no MPI transfer.
An example will be shown in Section~\ref{sec:load-balancing}.

\subsection{MPI parallel model}

$\lambda$PIC uses a patch-centric MPI model. Each MPI rank owns a subset of patches, and all intra-rank operations (field updates, particle pushes, current deposition) are performed by iterating over the local patch list.
Communication is required only for guard-cell synchronization, particle migration across patch boundaries, and dynamic load balancing.

After every Maxwell update and current deposition the guard cells of the field arrays must be refreshed.
Intra-rank neighbors are handled directly by copying the boundary layer of one patch into the guard cells of its neighbor.
Inter-rank synchronization uses non-blocking point-to-point communication with MPI derived datatypes.
For each boundary of each patch, a derived datatype extracts the interior boundary region on the sending side and deposits it directly into the guard cells on the receiving side.
The same pattern is used for current synchronization, with manual buffer packing for the four current-density arrays $J_x$, $J_y$, $J_z$, and $\rho$.

Particles that cross a patch boundary after the position push are transferred to the neighboring patch, and if that neighbor resides on a different MPI rank, an inter-rank transfer is invoked after the intra-rank transfer is complete.
The particle migration is implemented in C extension.
It proceeds in two phases. First, each rank counts how many particles are leaving through each boundary (face, edge, and corner), and these counts are exchanged with the neighbor patches.
The patch computes how many slots it must allocate (incoming particles minus dead particles) and pre-extends the particle arrays.
Then, the incoming particles fill the pre-allocated dead slots; outgoing particles are marked dead.
For inter-rank transfers, particle data are first packed into contiguous send buffers and sent to neighboring ranks with non-blocking send and receive.

During the simulation loop, each patch accumulates a load metric equal to its particle count plus half its cell count, and rebalancing is triggered once the imbalance across ranks exceeds a configurable threshold.
The rebalancing cycle begins by gathering per-patch loads and patch skeletons on rank~0, where skeletons are lightweight metadata that do not carry field or particle data.
A new partition is then computed from these loads.
The new assignment is broadcast to all ranks, after which ranks exchange the full patch objects for those patches that changed owner.
Finally, each rank rebuilds its local neighbor tables.
The threshold adapts automatically. It is relaxed by a factor of $\mathrm{e}/2$ following a failed rebalance, and tightened toward its initial value by a factor of $3/\pi$ following a successful one. The irrational factors are chosen to avoid resonance with periodic load fluctuations. This heuristic prevents oscillatory rebalancing in regimes where the particle distribution changes rapidly.

\section{Numerical Methods}\label{sec:numerical}

\subsection{PIC methods}\label{sec:pic-methods}

The particle push follows the standard Boris algorithm for relativistic charged particles, using the Cartesian momentum $\mathbf{u}=\gamma\mathbf{v}/c$ and the half-acceleration / rotation / half-acceleration sequence~
\cite{borisRelativisticPlasmaSimulationoptimization1970,c.k.birdsallPlasmaPhysicsComputer1991}.
Both two- and three-dimensional variants are implemented.
The position advance is a simple leapfrog step $x^{n+1/2}=x^{n-1/2}+\Delta tu^{n}/\gamma^{n}$, where $\gamma=\sqrt{1+u^{2}/c^{2}}$.

Field quantities are gathered from the Yee grid to particle positions with bilinear interpolation in 2D and trilinear interpolation in 3D.
The interpolation stencil respects the spatial staggering of the grid, so each field component is interpolated from its natural node locations.

Current deposition uses the Esirkepov charge-conserving scheme~\cite{esirkepovExactChargeConservation2001} with a quadratic (second-order) spline shape function.
The algorithm deposits the current density $\mathbf{J}$ and charge density $\rho$ by tracing each macro-particle's trajectory from $t-\Delta t/2$ to $t+\Delta t/2$ and accumulating the line integral along the path.
The resulting current automatically satisfies the discrete continuity equation on the Yee grid.
The implementation is written in C with OpenMP parallelization over patches; a fast-path variant precomputes the geometric factors when the charge and grid spacing are constant across a species.

\subsection{Space and time discretization}\label{sec:discretization}

Electromagnetic fields are discretized on a Yee staggered grid~
\cite{yeeNumericalSolutionInitial1966} with the standard leapfrog time integration with Courant--Friedrichs--Lewy (CFL) condition of $\Delta t=\eta_{\text{CFL}}/\bigl(c\sqrt{\Delta x^{-2}+\Delta y^{-2}+\Delta z^{-2}}\bigr)$.
The Maxwell equations
\begin{align}
    \frac{\partial\mathbf{E}}{\partial t} &= c^{2}\nabla\times\mathbf{B} - \frac{\mathbf{J}}{\varepsilon_{0}}, \label{eq:maxwell-e} \\
    \frac{\partial\mathbf{B}}{\partial t} &= -\nabla\times\mathbf{E}, \label{eq:maxwell-b}
\end{align}
are advanced in time as
\begin{align}
    \mathbf{E}^{n+1/2} &= \mathbf{E}^{n-1/2} + \Delta t\cdot\left(c^{2}\nabla\times\mathbf{B}^{n} - \frac{1}{\varepsilon_{0}}\mathbf{J}^{n}\right), \label{eq:maxwell-yee-e} \\
    \mathbf{B}^{n+1} &= \mathbf{B}^{n} - \Delta t\cdot\left(\nabla\times\mathbf{E}^{n+1/2}\right). \label{eq:maxwell-yee-b}
\end{align}
It should be noted that in the standard simulation loop of \texttt{Simulation} class, the time integration is split into two $\Delta t/2$ parts for consistent interpolation at the same time (see Section~\ref{sec:pic-loop}).

Convolutional Perfectly Matched Layer (CPML) absorbing boundaries~\cite{berengerPerfectlyMatchedLayer1994} is ported from the EPOCH code~\cite{arberContemporaryParticleincellApproach2015} and is attached to the boundary patches of the simulation box.

\subsection{PIC loop}\label{sec:pic-loop}

One complete simulation proceeds as Algorithm~\ref{alg:pic-timestep} (user callbacks may be attached at every stage; see Section~\ref{sec:architecture}).
The time-staggered leapfrog approach brings the $\mathbf{E}$ and $\mathbf{B}$ fields to the same temporal point when needed for the particle push, and makes it straightforward to include additional physics packages that require electromagnetic fields at either full- or half-step values~
\cite{arberContemporaryParticleincellApproach2015}.

\begin{algorithm*}[t]
    \caption{Main PIC simulation loop in $\lambda$PIC.}
    \label{alg:pic-timestep}
    \SetAlgoLined
    \SetKwComment{Comment}{\# }{}
    \SetKwFor{For}{for}{do}{end}
    \SetKwFor{ForEach}{for each}{do}{end}

    \Comment{Initialization}
    initialize patches, fields, MPI communicators, and solvers\\
    register species, pushers, and QED modules\\

    \BlankLine
    \Comment{Main time-stepping loop}
    \For{$\text{istep} = \text{itime}$ \KwTo $n_{\text{steps}}$}{

        \BlankLine
        \Comment{Field half-update $t \to t+\Delta t/2$}
        advance $E$ and $B$ by $\Delta t/2$\\
        synchronize guard cells across patches and MPI ranks\\

        \BlankLine
        \Comment{Particle loop (per species)}
        \ForEach{species}{
            sort particles by cell index\\
            push positions by $0.5\Delta t$\\
            interpolate $E$, $B$ to particle positions\\
            \uIf{has QED}{
                process nonlinear Compton and Breit--Wheeler events\\
            }
            push momenta by $\Delta t$ (Boris pusher)\\
            push positions by $0.5\Delta t$\\
            deposit current $J$\\
        }

        \BlankLine
        synchronize currents (intra-rank, then inter-rank)\\

        \BlankLine
        create photons and $e^+e^-$ pairs after all species pushed\\

        \BlankLine
        migrate particles across patch boundaries (intra-rank, then inter-rank)\\

        \BlankLine
        \uIf{load imbalance $>$ threshold}{
            perform dynamic load balancing (compute graph partition + MPI migration)\\
        }

        \BlankLine

        \BlankLine
        \Comment{Field half-update $t+\Delta t/2 \to t+\Delta t$}
        advance $B$ and $E$ by $\Delta t/2$\\
        synchronize guard cells across patches and MPI ranks\\

        \BlankLine
        advance simulation time $t \leftarrow t + \Delta t$\\
    }

    \BlankLine
\end{algorithm*}

User-defined callbacks can be attached at any stage in the main loop.
The public callback stages are defined at the beginning, start of the loop, between field updates, current deposition, QED particle creation, end of the loop and before exiting the simulation run.

It is worth noting that when no user callbacks are registered at the intermediate stages between position push, field interpolation, momentum push, and current deposition, $\lambda$PIC fuses these operations into a single loop over particles with a dedicated C kernel.
This reduces memory bandwidth pressure and eliminates the overhead of repeated Python-level dispatch.

\subsection{QED modules}\label{sec:qed}
The nonlinear Compton scattering module handles photon emission from high-energy electrons in strong electromagnetic fields.
For each macro-particle the quantum nonlinearity parameter $\chi_{e}=(\hbar/m_{e}c^{2})\sqrt{(F_{\mu\nu}p^{\nu})^{2}}$~\cite{ritusQuantumEffectsInteraction1985} is computed from the local field tensor and macro-particle momentum.
The emission probability is sampled from the spectrum under local constant field approximation; when an emission event occurs, the photon energy is drawn from the same distribution and the emitting electron's momentum is recoiled accordingly.
Emitted photons are buffered and added to the simulation after the species loop completes.

The Breit-Wheeler pair production module creates electron-positron pairs from high-energy photons.
The photon quantum parameter $\chi_{\gamma}$ is evaluated analogously to $\chi_{e}$; when a pair-production event is sampled, the energy is partitioned between the electron and positron according to the pair-energy distribution and the two new particles are inserted into the corresponding species arrays.

Both modules operate via an event-based Monte Carlo scheme~\cite{duclousMonteCarloCalculations2010,ridgersModellingGammarayPhoton2014}.
The optical depth is decremented each timestep and an event is triggered when it crosses zero.
Because QED cascades can create and destroy macro-particles at very high rates, these modules benefit directly from the lazy particle-deletion strategy described in Section~\ref{sec:class-hierarchy}.

\section{Code Verification and Benchmark}\label{sec:verification}

\subsection{Numerical heating}

A fundamental consistency check for any PIC code is the conservation of total energy in the absence of external drivers or dissipative boundaries~\cite{c.k.birdsallPlasmaPhysicsComputer1991}.
In a closed system the sum of particle kinetic energy and electromagnetic field energy should remain constant; any systematic drift is a measure of numerical heating or cooling introduced by the time and space discretization, a well-known artifact of PIC methods on a staggered grid~\cite{langdonEffectsSpatialGrid1970}.

\begin{figure}[]
  \centering
  \includegraphics[width=\columnwidth]{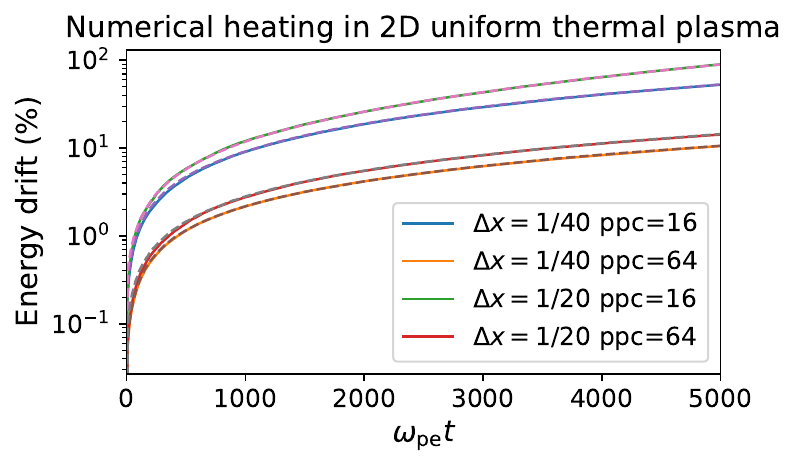}
  \caption{
    Normalized energy drift in a 2D uniform thermal plasma ($n_e = 10\,n_c$, $T = 1$~keV).
    Solid curves show $\lambda$PIC results for two spatial resolutions and two PPC counts; dashed curves show the corresponding EPOCH results.
  }
  \label{fig:numerical_heating}
\end{figure}

We initialize a uniform two-dimensional electron--ion plasma with density $n_e = 10\,n_c$ and temperature $k_\mathrm{B}T = 1$~keV in a square domain with periodic boundaries in both directions.
The Boris pusher~\cite{borisRelativisticPlasmaSimulationoptimization1970}, bilinear field interpolation and charge-conserving current deposition~\cite{esirkepovExactChargeConservation2001} are used with a CFL factor of $0.95$.
After initializing the Maxwell--J\"uttner momentum distribution for both species we evolve the system for $5000\,\omega_{pe}^{-1}$ and record the total energy every $10$ steps.
Fig.~\ref{fig:numerical_heating} shows the normalized energy drift
$\Delta E/E_0 = (E_\mathrm{kin}+E_\mathrm{field}-E_0)/E_0$
for two spatial resolutions ($\Delta x = \lambda_0/20$ and $\lambda_0/40$) and two PPC counts ($16$ and $64$).
These trends are consistent with the EPOCH results, confirming that the two codes exhibit the same numerical-heating behavior in this standard benchmark.

\subsection{Laser-plasma interaction}

To benchmark $\lambda$PIC in a highly nonlinear regime, we simulate an intense laser pulse interacting with a uniform overdense plasma in two dimensions and compare the results with EPOCH.
The laser has a wavelength $\lambda_0 = 0.8\,\mathrm{\mu m}$, a peak normalized vector potential $a_0 = 100$, a spot size $w_0 = 5\,\mathrm{\mu m}$, and a $\sin^2$ temporal envelope with a full duration $c\tau = 20\,\mathrm{\mu m}$.
It is injected from the left boundary into a uniform plasma ($e^-$ and $H^+$) of density $n_e = 10\,n_c$ that begins at $x = 5\,\mathrm{\mu m}$ and extends to the right-hand boundary.
The simulation domain is $25\,\mathrm{\mu m} \times 20\,\mathrm{\mu m}$, resolved with $\Delta x = \lambda_0/50$ and 10 macro-particles per cell per species.

\begin{figure}[]
  \centering
  \includegraphics[width=\columnwidth]{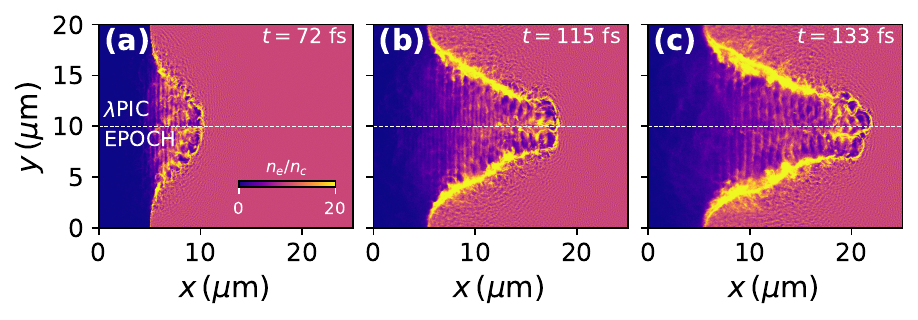}
  \caption{
    Electron density $n_e/n_c$ in a 2D laser-plasma interaction ($a_0 = 100$, $n_e = 10\,n_c$).
    The upper half of each panel shows the $\lambda$PIC result and the lower half shows the EPOCH result at $t = 72$, $115$, and $133$~fs.
  }
  \label{fig:hole_boring}
\end{figure}

Fig.~\ref{fig:hole_boring} shows the electron density $n_e/n_c$ at $t = 72$, $115$, and $133$~fs.
At the earliest time the laser front has pushed the plasma surface into a curved density spike, and by $t = 115$~fs a pronounced channel has formed with filamentary structure along its walls.
At $t = 133$~fs the channel has deepened and the transverse modulations have developed further.
The two codes reproduce the same density and filamentation patterns at each time step, confirming that $\lambda$PIC correctly captures the laser-plasma interactions in this regime.

\section{Scaling}\label{sec:scaling}

\subsection{Strong and weak scaling}\label{sec:scaling-results}

\begin{figure}[]
  \centering
  \includegraphics[width=\columnwidth]{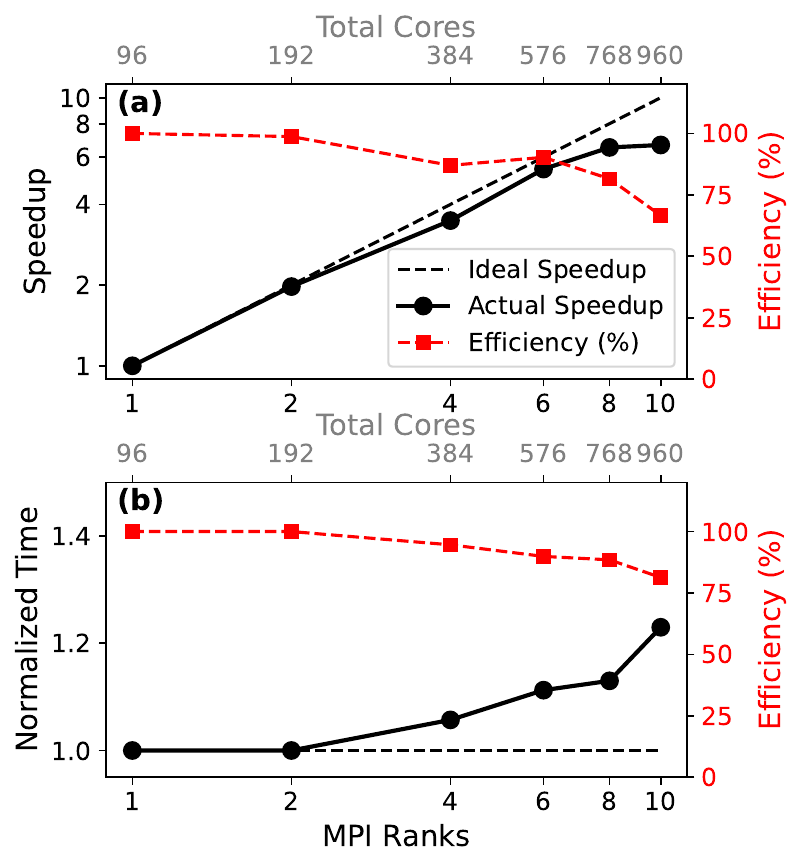}
  \caption{Strong (a) and weak (b) scaling. Black circles denote measured data, the black dashed line denotes ideal scaling, and red squares denote parallel efficiency (right-hand axis).}

  \label{fig:scaling}
\end{figure}

Fig.~\ref{fig:scaling} summarizes the parallel scaling of $\lambda$PIC on a two-dimensional Weibel-instability benchmark.
The benchmark problem consists of two counter-streaming electron beams drifting through a stationary background plasma with four species (beam~1, beam~2, background electrons, and ions) at $256$ macro-particles per cell per species.
Load balancing is disabled so that the scaling reflects only the communication overhead of field and particle exchanges.

These results demonstrate that the Python/Numba/C-extension hybrid in $\lambda$PIC delivers good parallel efficiency while retaining full runtime flexibility.
The Python layer is responsible for orchestration and callback dispatch, whereas all compute-intensive operations (particle push, current deposition, field update, and MPI communication) are executed in compiled C or Numba kernels.
Consequently, the performance of $\lambda$PIC is bounded by the same communication and surface-to-volume effects that limit pure compiled PIC codes, not by the Python interpreter.

For strong scaling the global grid is fixed at $1600\times1600$ cells, giving a total particle count of approximately $2.
6\times10^{9}$.
Each MPI rank uses $96$ OpenMP threads; the $1$- and $2$-rank cases are executed on a single node, whereas $4$ and more ranks span multiple nodes.
Strong scaling maintains parallel efficiency well above $80\%$ up to roughly $400$ cores, confirming that the Python orchestration layer does not become a bottleneck at this scale.
Beyond this point the curve gradually saturates, and the efficiency falls to approximately $70\%$ at the highest core count tested ($960$ cores).
The marked drop in efficiency at $4$ ranks ($384$ cores) coincides with the transition from intra-node to inter-node MPI communication, which carries a substantially higher latency than shared-memory exchanges within a node.
At larger core counts the additional slowdown is the familiar strong-scaling effect where the ratio of surface area (guard-cell and particle-migration volume) to volume (local compute) grows as the subdomain size shrinks.

For weak scaling the local grid size per MPI rank is held constant and the global problem size increases in proportion to the number of ranks.
The normalized elapsed time stays close to the ideal constant value up to a few hundred cores, and the parallel efficiency remains above $80\%$ across the entire range tested.
The gradual decline at higher core counts reflects the increasing cost of global reductions and load-imbalance noise as the total particle count grows, but the overall weak-scaling behavior is satisfactory for production runs.

\subsection{Single-node performance comparison}\label{sec:speed-comparison}

To assess the absolute performance of $\lambda$PIC against a mature compiled code, we compare the timestep throughput of $\lambda$PIC and EPOCH~\cite{arberContemporaryParticleincellApproach2015} on an identical two-dimensional uniform thermal plasma.
The benchmark problem uses a $1024\times1024$ grid with periodic boundaries, electron and proton species at $n_e = 10\,n_c$ and $k_\mathrm{B}T = 1$~keV, and $32$ macro-particles per cell per species, giving a total of approximately $6.7\times10^{7}$ particles.
No file output is performed during the timed loop, so the measurement reflects only particle push, field solve, and guard-cell exchange.
Both codes are run with $N$ MPI processes on $N$ cores, one thread per process, ensuring a direct comparison under the same parallelization model.
The benchmark is run on a single socket with $96$ cores.

\begin{figure}[]
  \centering
  \includegraphics[width=\columnwidth]{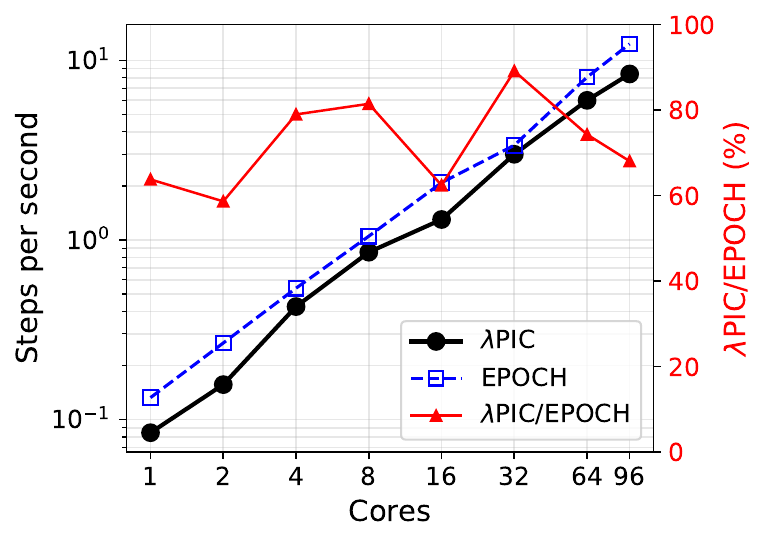}
  \caption{
    Timestep throughput (left axis) and $\lambda$PIC-to-EPOCH throughput ratio (right axis) on a 2D uniform thermal plasma ($1024\times1024$ grid, $32$~particles per cell per species, $n_e = 10\,n_c$, $T = 1$~keV, periodic boundaries).
    Both codes use $N$ MPI processes with one thread each on a single $96$-core socket.
  }
  \label{fig:speed_comparison}
\end{figure}

Fig.~\ref{fig:speed_comparison} shows the timestep throughput and the $\lambda$PIC-to-EPOCH ratio versus core count from $1$ to $96$ cores.
Both codes scale close to ideally up to $96$ cores, confirming that the Python orchestration layer does not become a scaling bottleneck.
The $\lambda$PIC-to-EPOCH ratio fluctuates between approximately $60$ and $90\%$ across the range, indicating that $\lambda$PIC achieves a substantial fraction of EPOCH's performance without resorting to a monolithic compiled core.
The residual gap reflects the maturity of EPOCH's hand-tuned Fortran kernels; the particle pusher and the cross-rank boundary particle exchange in $\lambda$PIC still have room for optimization, particularly in reducing communication overhead at high core counts.
Nonetheless, the results confirm that the Python/Numba/C-extension hybrid incurs only a modest performance penalty relative to a monolithic compiled code while retaining the flexibility of the callback-centric architecture.

\subsection{Dynamic load balancing}\label{sec:load-balancing}

\begin{figure*}[]
  \centering
  \includegraphics[width=\textwidth]{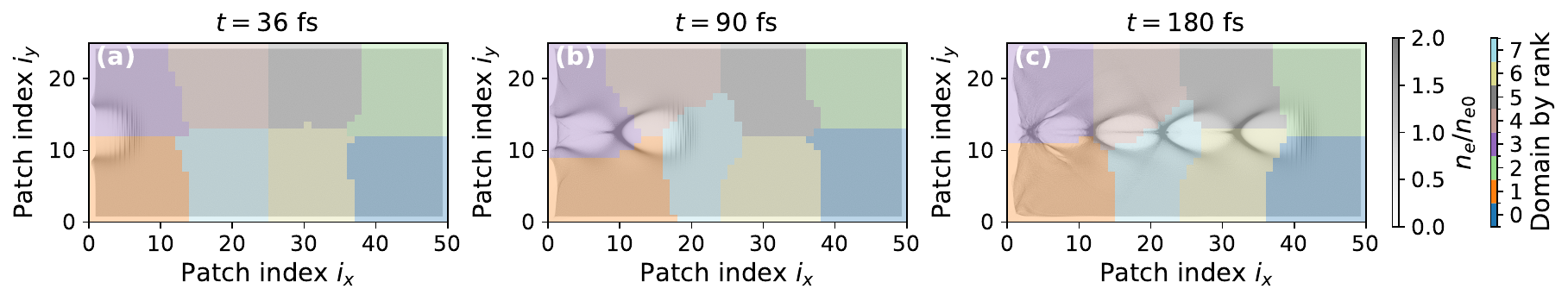}
  \caption{
    Dynamic load balancing of a laser-wakefield acceleration simulation with 8 MPI ranks.
    Electron density $n_e/n_{e0}$ is shown in grayscale, with ranks shown in different colors. 
    (a)~$t=36$~fs, (b)~$t=90$~fs, (c)~$t=180$~fs.
  }
  \label{fig:load-balance-partitions} 
\end{figure*}

The dynamic load balancer described in Section~\ref{sec:architecture} redistributes patches when the per-rank imbalance exceeds an adaptive threshold, using a graph partitioner (METIS~\cite{karypisFastHighQuality1998} in this implementation).
Fig.~\ref{fig:load-balance-partitions} illustrates the resulting spatial reorganization for a two-dimensional laser-wakefield acceleration problem.
A laser pulse with $\lambda=0.8~\mu$m, $a_0=2$ and $w_0=4~\mu$m enters from the left into a uniform electron plasma of density $n_e=0.005\,n_c$.
The domain is $60~\mu$m~$\times$~$30~\mu$m and is divided into $50\times25$ patches distributed over eight MPI ranks.
As the laser propagates and transforms the electron distribution, patches are redistributed to ensure even computational load across ranks.
It is worth noting that because we maximize the overlap of patch sets between the old and new distributions, the set of patches held by any individual rank does not change drastically.
The graph partitioner itself offers no such guarantee, because it does not take the historical partition into account.

\section{Application Examples}\label{sec:applications}

Typical callback applications include in-situ reduced diagnostics such as particle energy spectra and angular distributions, selective particle tracing based on momentum or quantum nonlinearity thresholds, dynamic particle injection and on-the-fly synchrotron radiation spectra.
These use cases are conceptually straightforward and can be implemented with a few lines of Python.
Below we turn to three more elaborate examples that couple multiple stages and alter the simulation domain geometry.

\subsection{Species-resolved electromagnetic fields}

In a conventional PIC simulation the Maxwell solver evolves the total electromagnetic field produced by all species collectively.
While this is sufficient for the self-consistent dynamics of the plasma, there are situations in which one wishes to know the field generated by a single species.
Such separation is physically meaningful since Maxwell equations are linear and the per-species solutions reflect the real radiation from each species.
Retrieving the species-resolved fields requires solving the Maxwell equations (Eqs.~\ref{eq:maxwell-e}--\ref{eq:maxwell-b}) with only the current density of interest as the source term.
The per-species time evolution is integrated with the leap-frog scheme (Eqs.~\ref{eq:maxwell-yee-e}--\ref{eq:maxwell-yee-b}).
This requires either dumping the current density at each time step or restructuring the source code for per-species Maxwell solvers.
In the $\lambda$PIC framework, this is simplified with the callback mechanism.
Because $\lambda$PIC exposes the current deposition stage to callbacks, one can copy the per-species current after current deposition and advance the Maxwell equations on-the-fly.

\begin{figure}[]
  \centering
  \includegraphics[width=\columnwidth]{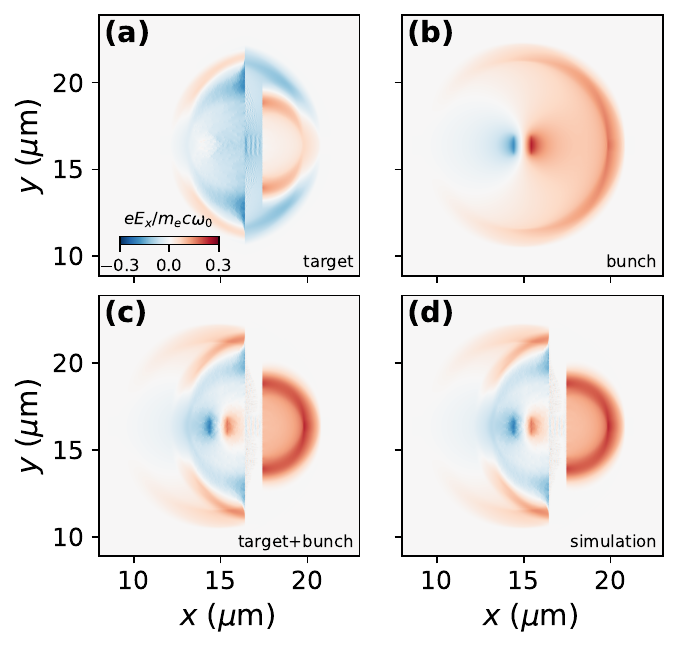}
  \caption{
    Species-resolved reconstruction of the longitudinal electric field for a relativistic electron bunch traversing an overdense plasma slab.
    (a) $E_x$ driven by the plasma-slab currents alone.
    (b) $E_x$ driven by the bunch currents alone.
    (c) Sum of (a) and (b).
    (d) Total $E_x$ from the built-in Maxwell solver.
  }
  \label{fig:species-resolved}
\end{figure}

For example, when analyzing the driven source of the transition-radiation-enhanced proton acceleration~\cite{qinTransitionRadiationField2026}, the fields generated by the electron bunch and the target are extracted with $\lambda$PIC.
Fig.~\ref{fig:species-resolved} shows the results of a relativistic electron bunch ($n_b = 0.1n_c$, $u_x = 100$, width $1\,\mu\text{m}$) traversing a plasma slab ($n_e = 10n_c$, thickness $1\,\mu\text{m}$).
Figs.~\ref{fig:species-resolved}(a) and (b) show the longitudinal electric field driven by the plasma slab and the electron-bunch currents, respectively.
The close agreement between (c) and (d) confirms the correctness of this method.
The results reveal the fact that the transition radiation is not the field generated by the bunch or the target alone, but both.
The bunch generates a dipole field and the target reacts to the bunch field, generating fields that modify the former, producing a standard transition radiation field on the rear of the target~\cite{carronFieldsParticlesBeams2000}.
This example illustrates how the callback-centric design allows arbitrary physics modules to be injected into the timestep pipeline without modifying the core solver.

\subsection{Non-rectangular domain}\label{sec:annular}

Traditional PIC frameworks decompose the simulation box into a regular Cartesian array of rectangular patches.
While non-rectangular domain is possible with patch-based PIC frameworks, maintaining load balance across MPI ranks is difficult with space-filling curves.
Graph-based domain decomposition removes this restriction because patches are treated as vertices in an adjacency graph rather than as fixed patches in a rectangular array.
Any connected subset of patches can be partitioned and balanced independently of the global grid geometry, enabling efficient dynamic load balancing even for irregular domains.
This capability is particularly valuable for irregular simulation domains that arise in studies of laser-plasma interactions and advanced accelerator concepts~\cite{chenTunableSynchrotronlikeRadiation2016, luoMultistageCouplingLaserWakefield2018}.

\begin{figure}[]
  \centering
  \includegraphics[width=\columnwidth]{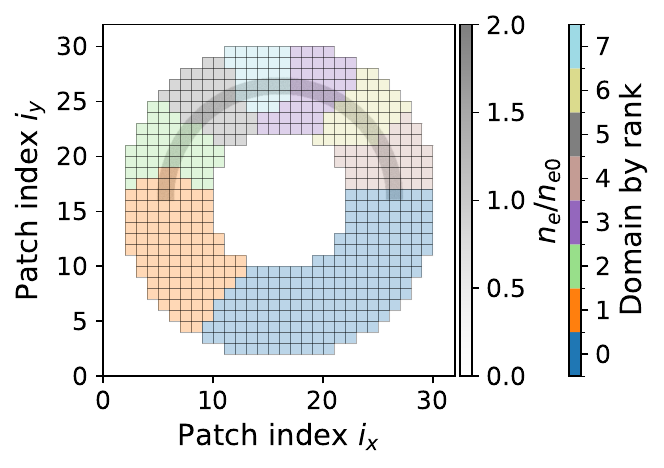}
  \caption{
    Annular simulation domain created with a mask function on 8 ranks.
    The domain patches are colored by MPI ranks.
    Electron density (grayscale) is a half-ring in the upper domain for demonstration of load-balancing.
  }
  \label{fig:ring-domain}
\end{figure}

To demonstrate this capability we construct an annular (ring-shaped) domain defined by an inner radius $r_{\text{inner}} = 0.2 L_x$ and an outer radius $r_{\text{outer}} = 0.45 L_x$, centered in a $512\times 512$ cell box of size $L_x \times L_y$.
A boolean mask function evaluates whether each patch center lies inside the annulus; only patches that satisfy the mask are created.
For demonstration, all open boundaries are attached with PML layers for simplicity.
To emphasize the load-balancing capability, a half-ring plasma is loaded in the annular domain, leaving the lower half as vacuum.
The peak density is $n_e = 10n_c$ with ten macro-particles per cell for electrons and two for protons.
Fig.~\ref{fig:ring-domain} shows the resulting patch layout colored by 8 MPI ranks.
Because the particle load is concentrated in the upper half of the ring, ranks that own patches in that region receive a smaller angular wedge, while ranks in the lower vacuum region are assigned a larger wedge.
This example demonstrates the possibility to simulate in non-rectangular domain with graph-based decomposition.

\subsection{Hybrid fluid-PIC coupling}

Dense background plasmas are expensive to resolve with fully kinetic PIC because the particle count scales with the plasma density.
When the background is cold and the dynamics of interest are carried by a kinetic species, treating the background as a fluid while retaining a kinetic description for the beam is both physically sound and efficient.
Hybrid models of this kind are widely used to study high-current electron-beam transport in dense plasma targets~\cite{caoTwoDimensionalHybridModel2014} and fast-electron transport in solid-density targets~\cite{yangHybridPICFluid2023}, and general kinetic-ion/fluid-electron PIC algorithms have been implemented in open-source frameworks~\cite{leHybridVPICOpensourceKinetic2023}.

In the hybrid model used here the background electrons are described by the continuity equation
$\partial_t n_e + \nabla \cdot (n_e \mathbf{v}_e) = 0$,
with the fluid current density given by $\mathbf{J}_e = -e n_e \mathbf{v}_e$.
The fluid momentum is advanced with the relativistic Boris pusher, and the density is updated with an upwind continuity solver.
For demonstration we use this simple fluid model; more elaborate closures and higher-order schemes can be substituted through the same callback mechanism.
The fluid current is calculated after current deposition and added to the total current density that drives the Maxwell equations.

\begin{figure}[]
  \centering
  \includegraphics[width=\columnwidth]{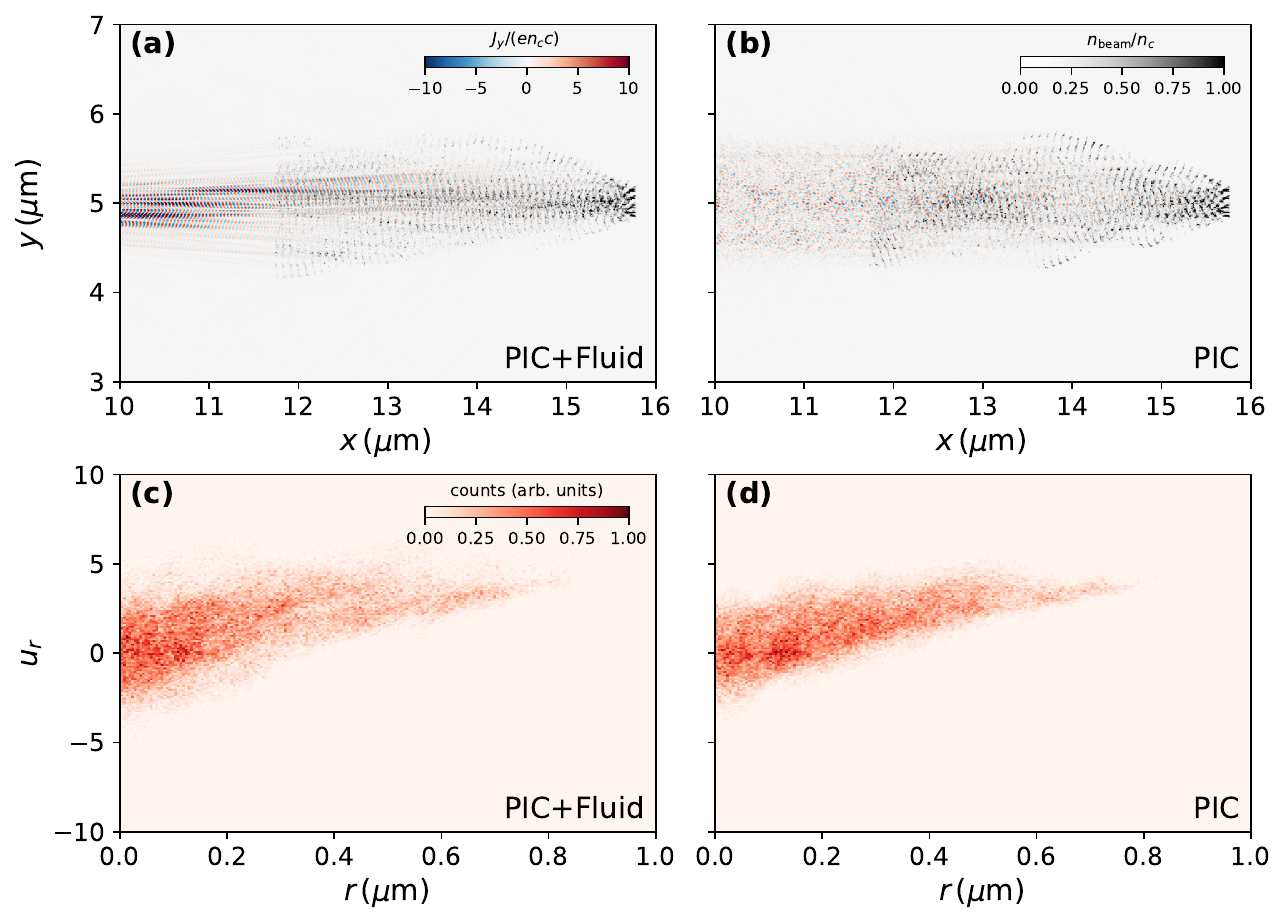}
  \caption{
    Relativistic electron beam transport in a high-density plasma.
    (a) Transverse current $J_y$ with beam density $n_{\text{beam}}$ for the hybrid fluid-PIC simulation and 
    (b) for the full-PIC simulation.
    (c) Transverse phase space $(r, \mathbf{u}\cdot\hat{\mathbf{r}})$ of the beam electrons for the hybrid fluid-PIC simulation and
    (d) for the full-PIC simulation.
  }
  \label{fig:hybrid-fluid-beam}
\end{figure}

As an example, we simulate the transport of a relativistic electron beam of density $n_b = 0.5n_c$, normalized momentum $u_x = 50$, and width $0.2\,\mu\text{m}$ propagating through a uniform plasma of density $n_e = 100n_c$ in 2D.
The background electrons are treated as a cold fluid, while the beam electrons and the background ions remain fully kinetic.
Fig.~\ref{fig:hybrid-fluid-beam} compares the hybrid description with a fully kinetic PIC reference.
Figs.~\ref{fig:hybrid-fluid-beam}(a) and (b) show the longitudinal current $J_y$ overlaid with the beam electron density.
The hybrid simulation reproduces the filamentary current pattern and the beam density modulation seen in the full-PIC reference at a qualitative level.
Figs.~\ref{fig:hybrid-fluid-beam}(c) and (d) show the transverse phase space $(r, u_r)$ of the beam electrons, where $u_r = \mathbf{u}\cdot\hat{\mathbf{r}}$ is the radial component of the normalized momentum.
The two descriptions are qualitatively consistent, confirming that the cold-fluid background approximation is adequate for capturing the main features of the beam dynamics in this regime.

This example demonstrates that a self-consistent fluid species can be coupled to the electromagnetic PIC solver without modifying the core timestep loop.
Unlike the species-resolved fields example, where the callback only reads the deposited current to compute an auxiliary field, the hybrid-fluid callback writes a current back into the Maxwell solver, creating a two-way coupling between the fluid and the kinetic species.
The hybrid model is implemented as a callback that operates on the field and current arrays exposed by the framework.

\section{Conclusions}\label{sec:conclusions}

$\lambda$PIC has resolved the tension between execution speed and extensibility in PIC software by making user-injected logic the primary extension mechanism.
Its callback-centric architecture has exposed every stage of the timestep loop to user-defined Python functions, while a Python/Numba/C-extension hybrid guarantees single-node performance and scaling efficiency comparable to monolithic compiled codes.
The callback mechanism has further enabled a broad class of user-defined physics modules, from custom diagnostics to multi-physics coupling, without modifying the core algorithms.
Graph-based domain decomposition has further enabled dynamic load balancing and non-rectangular geometries that are difficult to achieve with traditional Cartesian patch decompositions.
The code has been verified against standard benchmarks with direct comparison to EPOCH, with parallel scaling demonstrated up to $960$ cores.
$\lambda$PIC is open-source and publicly available at \url{https://github.com/xsgeng/lambdapic}; future directions include GPU acceleration and additional physics modules such as implicit solvers and nuclear physics.

\section*{Declaration of competing interest}
The authors declare that they have no known competing financial interests or personal relationships that could have appeared to influence the work reported in this paper.

\section*{Declaration of generative AI and AI-assisted technologies in the manuscript preparation process}
During the preparation of this work, the authors used OpenCode with Kimi-k2.5/k2.6 and GLM-5.2 for agentic coding, writing and visualization.
The authors reviewed and edited the output as needed and take full responsibility for the content of the published article.

\section*{Data availability}
The $\lambda$PIC source code is publicly available at \url{https://github.com/xsgeng/lambdapic} under the GNU General Public License v3.0 (GPL-3.0).

\section*{Acknowledgements}
This work is supported by the National Natural Science Foundation of China (Nos. 12388102, 12374298 and 12304384) and the Strategic Priority Research Program of Chinese Academy of Sciences (No. XDB0890303).

\printcredits

\bibliographystyle{cas-model2-names}
\bibliography{refs}

\end{document}